\documentclass[a4paper,11pt]{article}
\usepackage{pos}
\usepackage{amsmath,amsfonts}
\usepackage{bm}
\usepackage{aas_macros}

\newcommand{\dd}[0]{\ensuremath{\mathrm{d}}}
\renewcommand{\vec}[1]{\ensuremath{\bm{#1}}}
\newcommand{\bb}{\hat{\bm{b}}}

\title{Spectrum and anisotropies of Galactic cosmic rays: a laboratory for magnetic fields}
\ShortTitle{Spectrum and anisotropies of Galactic cosmic rays}

\author*[a]{Philipp Mertsch}

\affiliation[a]{Institute for Theoretical Particle Physics and Cosmology (TTK), RWTH Aachen University, \\ 
Sommerfeldstr.\ 14, 52074 Aachen, Germany}

\emailAdd{pmertsch@physik.rwth-aachen.de}

\abstract{
Much has been learned about Galactic cosmic rays in the past decade: 
On the observational side, the spectra of cosmic ray nuclei have been directly measured with high precision, resolving chemical composition up to TV rigidities. 
At even higher rigidities, direct detection is making contact with indirect observations from air shower arrays. 
A number of breaks have been found in the nuclear spectrum, which was previously thought to be a pure power law up to the knee. 
Data from air shower arrays also show interesting features in the arrival directions of cosmic-ray nuclei. 
On the theoretical side, more sophisticated models are able to explain the various spectral breaks either with transitions between different classes of sources or with changes in the transport regime. 
Yet, it has become clear that our ignorance of the structure of the Galactic magnetic fields, both on large and small scales, is limiting precision predictions. 
Turning this problem into an opportunity though, we can use Galactic cosmic rays as a laboratory for the study of Galactic magnetic fields. 
In this review talk, delivered at the 39th International Cosmic Ray Conference (ICRC2025), I have summarised what is known about the spectrum and anisotropies of Galactic cosmic rays, what is not known yet and what can be learnt in the future.
}

\ConferenceLogo{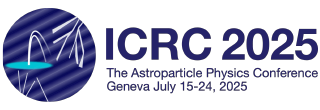}

\FullConference{39th International Cosmic Ray Conference (ICRC2025)\\
 15-24 July 2025\\
Geneva, Switzerland\\}

\begin{document}
\maketitle

\section{Introduction}

\begin{figure}[!t]
\includegraphics[]{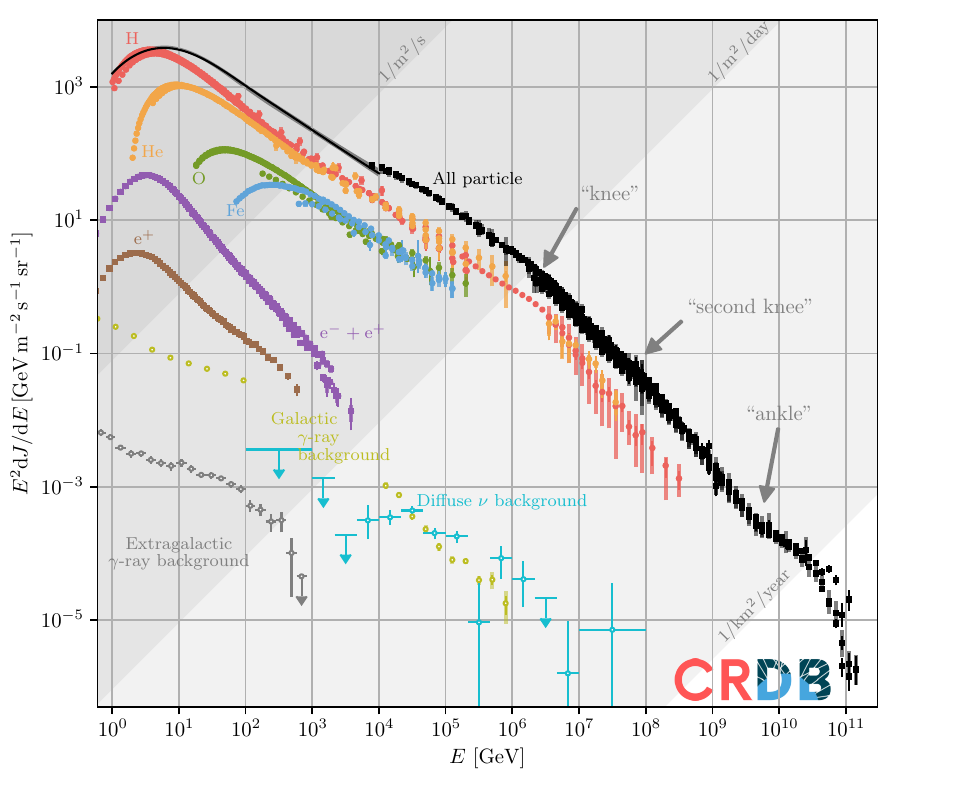}
\caption{
    The differential intensity $\dd J / \dd E$ of cosmic rays as a function of kinetic energy $E$. 
    The data on charged cosmic rays have been retrieved from the Cosmic-Ray Database (CRDB)~\cite{Maurin:2023alp} and from Refs.~\cite{LHAASO:2024knt}. 
    Data on gamma-rays and neutrinos are from Refs.~\cite{Fermi-LAT:2012edv,Fermi-LAT:2014ryh,LHAASO:2023gne,IceCube:2023qpn}. 
}
\label{fig1}
\end{figure}

The last one and a half decades half brought rapid progress to the study of Galactic cosmic rays (GCRs) thanks to improved direct measurements at GV-TV rigidities and high-statistics air-shower observations at TeV-PeV energies. 
On top of the spectral features identified in the all-particle spectrum decades ago, direct instruments have established multiple spectral features in proton, helium, and heavier nuclei; indirect measurements increasingly constrain composition and anisotropy at higher energies. 
For this year's ICRC, I was tasked with reviewing spectra and anisotropy of GCRs. 
This is a vast field that touches upon a number of other lines of research; fortunately, many of them were covered by colleagues in dedicated plenary talks: 
particular classes of sources, like supernova remnants~\cite{Lemoine-Goumard:2025}, pulsar wind nebulae~\cite{Recchia:2025}, star cluster winds~\cite{Peron:2025} or microquasars~\cite{Khangulyan:2025}; connections to $\gamma$-rays and neutrinos~\cite{Semikoz:2025}; and particle acceleration~\cite{Lemoine:2025}. 

A central theme in interpreting these data is the role of magnetic fields across scales. 
Magnetic turbulence and ordered fields regulate GCR transport and thereby shape spectral features and angular distributions. 
Conversely, GCR data offer diagnostics of interstellar magnetism, providing valuable insight into the strength and orientation of the ordered component; as well as the power spectrum and anisotropies of the turbulent component. 
In this review, I will adopt magnetic fields across scales as the organising principle that bridges spectra and anisotropies.

I will first summarise salient spectral measurements (emphasising differences and similarities between primary and secondary species), then discuss interpretations of the observed spectral breaks, subsequently invoking source populations, individual nearby sources, intrinsic source features and transport effects. 
I will briefly review data on electrons and positrons and their interpretation. 
Finally, I will outline constraints from the large-scale anisotropy of GCRs and lay out our current thinking about small-scale anisotropies. 
I will conclude with a brief summary.

\section{Spectrum: data}

\begin{figure}[b]
\includegraphics[width=\textwidth]{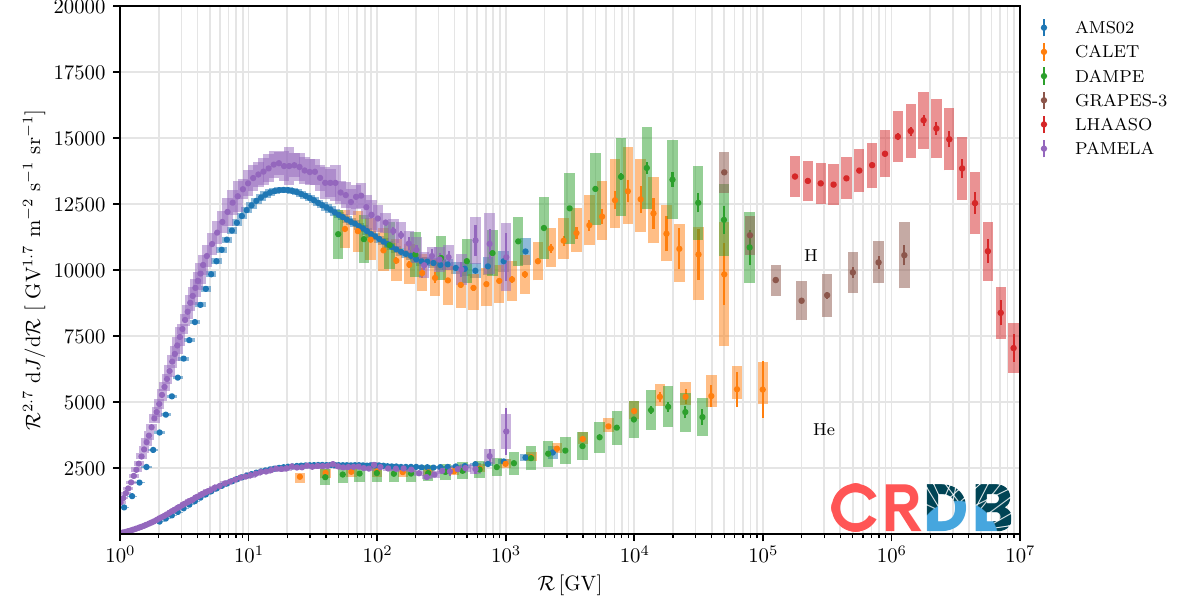}
\caption{
    The proton and Helium rigidity spectra between a GV and $\sim 10 \, \text{PV}$. 
    The intensity $\dd J / \dd \mathcal{R}$ has been scaled with $\mathcal{R}^{2.7}$ to bring out the various spectral features. 
    The data have been retrieved from the Cosmic-Ray Database (CRDB)~\cite{Maurin:2023alp} and from Refs.~\cite{GRAPES-3:2024mhy,LHAASO:2025byy}. 
}
\label{fig2}
\end{figure}

On the all-particle level, the CR intensity across $\sim 12$ orders of magnitude in energy follows a broken power law with the canonical ``knee'' at a few PeV and the ``ankle'' at a few EeV, see Fig.~\ref{fig1}. 
The data on proton and Helium from the space-based experiments AMS-02~\cite{AMS:2021nhj}, CALET~\cite{CALET:2022vro}, DAMPE~\cite{DAMPE:2019gys} and PAMELA~\cite{PAMELA:2010kea} as well as the air-shower arrays GRAPES-3~\cite{GRAPES-3:2024mhy} and LHAASO~\cite{LHAASO:2025byy} is shown in more detail in Fig.~\ref{fig2}. 
The proton spectrum in rigidity $\mathcal{R}$ is $\propto \mathcal{R}^{-2.8}$ above $\sim 10 \, \text{GV}$. 
At a few hundred GV, the spectrum gradually hardens, before it softens again around $\sim 10 \, \text{TV}$. 
Data from ground-based experiments indicate another hardening at a few hundred TV (already indicated in the combined $\text{p}+\text{He}$ spectrum measured by DAMPE~\cite{DAMPE:2023pjt}, not shown in Fig.~\ref{fig2}). 
LHAASO locates the proton knee at $3 \, \text{PV}$. 
The Helium spectrum is overall harder than the proton spectrum by a factor $\sim\mathcal{R}^{0.1}$, but the positions in rigidity of the first hardening and the first softening are the same as for proton. 
If only magnetic fields shaped the particle populations, then the rigidity spectra of different species should be the same. 

\begin{figure}
\includegraphics[width=\textwidth]{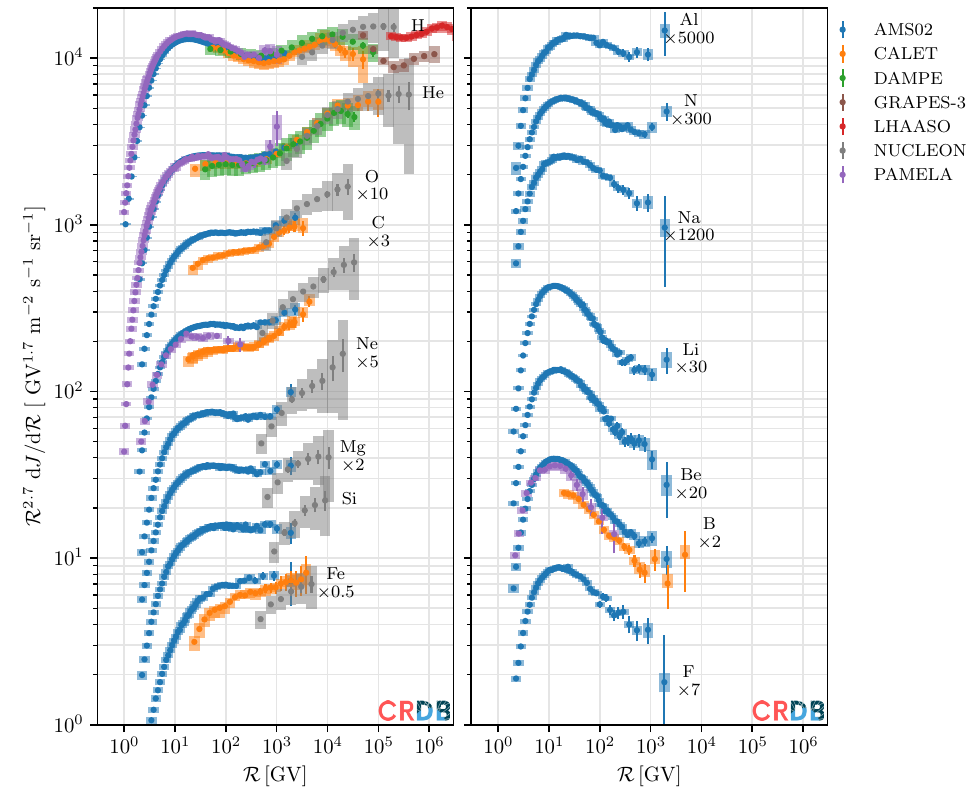}
\caption{
    The spectra of primary (left panel) and secondary GCRs (right panel). 
    The intensity $\dd J / \dd \mathcal{R}$ has been scaled with $\mathcal{R}^{2.7}$ to bring out the various spectral features. 
    The data have been retrieved from the Cosmic-Ray Database (CRDB)~\cite{Maurin:2023alp} and from Refs.~\cite{GRAPES-3:2024mhy,LHAASO:2025byy}. 
}
\label{fig3}
\end{figure}

The spectral universality is however realised for a number of species heavier than protons, see Fig.~\ref{fig3}. 
In fact, elemental spectra can be classified into two groups: primaries, e.g.\ H, He, O, C, Ne, Mg, Si and Fe, that are present and accelerated at source (left panel in Fig.~\ref{fig3}); and secondaries with strongly suppressed source abundances, e.g.\ Al, N, Na, Li, Be, B and F. 
The primary spectra largely resemble the Helium spectrum in shape. 
The secondary spectra are softer by a factor $\mathcal{R}^{0.4 \mathellipsis 0.5}$, which is expected from GCR transport, see below. 
Remarkably though, a spectral hardening is observed for almost all species around the same rigidity as for p and He. 
We note in passing that there are still experimental discrepancies, mostly in normalisation though. 

To go beyond qualitative statements, we will in the following adopt the standard transport formalism~\cite{Ginzburg1964}.

\section{Spectrum: interpretation}

The general GCR transport equation for species $j$ is a partial differential equation in space and momentum, including spatial diffusion and advection, momentum-space diffusion (reacceleration), continuous losses and gains, and catastrophic processes (decays, spallation), together with sources. 

Let's simplify the problem by constraining ourselves to the most important spatial variation in the vertical direction $z$, choosing to ignore the extent of the Galaxy in the radial direction altogether. 
For simplicity, we demand that diffusion be isotropic such that the diffusion \emph{tensor} reduces to a diffusion \emph{coefficient}. 
Furthermore, we ignore spatial dependencies in the diffusion coefficient and advection velocity. 
The transport equation then simplifies to~\cite{Ginzburg1964,Putze:2010zn} 
\begin{equation}
    \frac{\partial \psi_j}{\partial t} 
    - \kappa \frac{\partial^2 \psi_j}{\partial z^2} 
    + U \frac{\partial \psi_j}{\partial z} 
    + \left( v n_{\text{gas}}(z) \sigma_j 
    + \frac{1}{\tau_j} \right) \psi_j 
    = q_j(z) 
    + \sum_{j < k} \left( v n_{\text{gas}}(z) \sigma_{k \to j} 
    + \frac{1}{\tau_{k \to j}} \right) \psi_k \, . 
\end{equation}
If we further look for a steady state and approximate the gas and source distribution of height $h$ as infinitely thin, we find 
\begin{align}
    - \kappa \frac{\partial^2 \psi_j}{\partial z^2} 
    + U \frac{\partial \psi_j}{\partial z} 
    + \left( 2 h \delta(z) n_{\text{gas}}(0) v \sigma_j 
    + \frac{1}{\tau_j} \right) \psi_j 
    &= 2 h \delta(z) q_j(0) \nonumber \\ 
    & + \sum_{j < k} \left( 2 h \delta(z) n_{\text{gas}}(0) v \sigma_{k \to j} 
    + \frac{1}{\tau_{k \to j}} \right) \psi_k \, . 
\end{align}
This ordinary differential equation can be solved, taking into account the boundary conditions $\dd \psi_j / \dd z = 0$ for $z=0$ and $\psi_j = 0$ for $z = \pm L$, $L$ being the half-height of the propagation volume. 
Finally, to simplify further, we assume no advection and ignore decays. 
The analytical solution then takes the simple form, 
\begin{equation}
    \psi_j(0, \mathcal{R}) = \frac{h \left( q_j(\mathcal{R}) + \sum_{k>j} v n_{\text{gas}}(0) \sigma_{k \to j} \psi_k \right) }{\left( h v n_{\text{gas}}(0) \sigma_j + \dfrac{\kappa(\mathcal{R})}{z_{\text{max}}} \right) } \, , \label{eqn:propagated_spectrum}
\end{equation}
$q_j(\mathcal{R}) \equiv q_j(z=0, \mathcal{R})$ denoting the source spectrum. 
For standard parameter values, the spectral shape above a few GV will be determined by the source term in the numerator, $q_j(\mathcal{R})$, and the diffusion coefficient in the denominator, $\kappa(\mathcal{R})$. 
If we adopt a power-law for the source spectrum, as predicted by diffusive shock acceleration~\cite{1977DoSSR.234.1306K,1977ICRC...11..132A,Blandford:1978ky,Bell:1978zc}, e.g.\ $q_j(\mathcal{R}) \propto \mathcal{R}^{-\gamma}$ and a power-law diffusion coefficient, $\kappa(\mathcal{R}) \propto \mathcal{R}^{\delta}$, we recover a power-law GCR spectrum, that is softer than the source spectrum, $\psi_j(\mathcal{R}) \propto \mathcal{R}^{-\gamma-\delta}$. 
As a numerical example, if the source spectrum were slightly softer than the canonical $\gamma = 2$, e.g.\ $\gamma = 2.2$ and if the diffusion coefficient had the rigidity dependence given by $\delta = 0.6$, this could explain the locally observed proton spectrum, $\psi_{\text{p}} \propto \mathcal{R}^{-2.8}$. 

In order to explain deviations from a single power law, it seems we can either introduce such features in the numerator, that is modify the source spectrum, or in the denominator, that is modify the transport. 
Here we consider four classes of effects that need not be mutually exclusive.

\subsection{Multiple source populations}

\begin{figure}
    \centering
\includegraphics[scale=0.8]{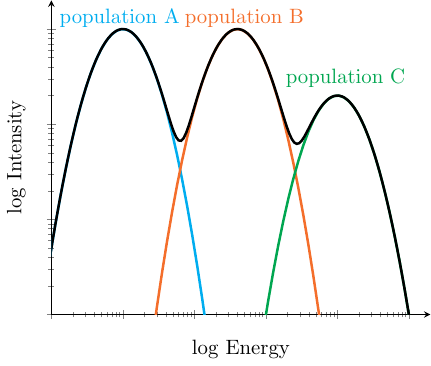}
\includegraphics[scale=0.8]{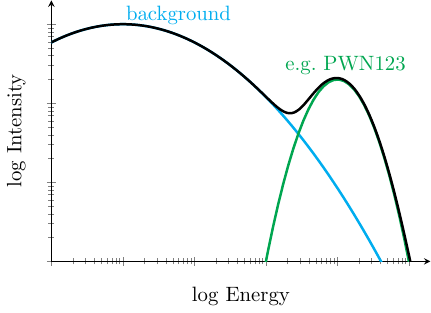}
\caption{\textbf{Left:} Illustration of the superposition of the contributions from different source classes, generically labelled `A', `B' and `C'. \textbf{Right:} Superposition of the spectrum from a background population and one individual source, in this example a pulsar wind nebulae.}
\label{fig4}
\end{figure}

Different accelerator classes (or environments) naturally superpose to produce bumps and smooth features if their individual spectra are hard enough and their maximum rigidities differ. 
The resulting spectrum can contain as many bumps and wiggles as there are source classes; see the left panel of Fig.~\ref{fig4} for an illustration. 
Within a supernova-remnant (SNR) paradigm alone, a range of explosion energies and circumstellar environments can lead to effective sub-populations with different cut-offs, e.g.~\cite{Zirakashvili:2015mua,Cristofari:2020mdf}. 
Variants invoke a low-energy contribution from isolated SNRs and a higher-energy contribution from SNe exploding in stellar clusters or superbubbles, whose winds and collective shocks can plausibly reach higher maximum rigidities, e.g.~\cite{Vieu:2022wsc}. 
Such superpositions can account for a gradual hardening and a softening feature near $\sim 10\,\mathrm{TV}$, without fine-tuning beyond physically reasonable ranges.

\subsection{Individual nearby sources}

A distinct high-energy bump or wiggle may result from an individual, relatively nearby, young, and powerful source (e.g., a SNR or pulsar/SNR complex), superimposed on a smooth Galactic background, e.g.~\cite{Erlykin:2012zz,Kachelriess:2018ser,Fornieri:2020kch} and many more; this is illustrated in the right panel of Fig.~\ref{fig4}. 
However, probabilistic considerations are essential: the number of sources contributing at rigidity $\mathcal{R}$ scales with the source rate times an effective residence time $t_{\mathrm{esc}}$ and the area of the diffusive ``horizon''. 
Denoting the halo half-height by $L$ and the diffusion coefficient by $\kappa(R)$, one estimates $t_{\mathrm{esc}} \sim L^2/\kappa(R)$ and a horizon $r_{\mathrm{hor}}\sim \sqrt{4 \kappa(R) t_{\mathrm{esc}}}\sim L$. For sources with a rate similar to galactic supernovae this implies thousands to hundreds of thousands of contributing sources at $\mathcal{O}(10 \, \mathrm{GV})$. 
This number decreases to $\mathcal{O}(10^2)$ at $10\,\mathrm{TV}$, and to potentially only a handful near the knee. 
Therefore at least the spectral feature at hundreds of GV is unlikely due to individual nearby sources~\cite{Genolini:2016hte}. 

This constraint is weakened if individual sources produced narrow spectral features. 
This is realised rather naturally if rigidity-dependent source escape (as implied by models of magnetic field amplification during shock acceleration) and stochastic source modelling are combined~\cite{Stall:2024klf}. 
Remarkably, both ingredients are standard fare in the supernova paradigm for GCRs that requires efficient amplification of magnetic fields in the shock waves of SNRs; therefore, observable features are to be expected. 
Turning this around, spectral features can be employed as a tool for testing models of magnetic field amplification.

\subsection{Intrinsic source spectral features}

\begin{figure}
    \begin{tabular}{l p{0.1cm} c p{0.1cm} c}
        Scenario & & Primaries & & Secondaries \\
        \hline \\[-1em]
        Break in source spectrum & & \raisebox{-0.5\totalheight}{\includegraphics[]{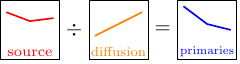}} & & \raisebox{-0.5\totalheight}{\includegraphics[]{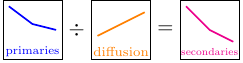}} \\
        Break in diffusion coefficient & & \raisebox{-0.5\totalheight}{\includegraphics[]{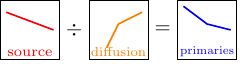}} & & \raisebox{-0.5\totalheight}{\includegraphics[]{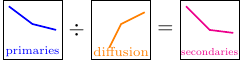}} 
    \end{tabular}
    \caption{Formation of breaks in primary and secondary spectra, either due to a break in the source spectrum (top) or a break in the diffusion coefficient (bottom)}
    \label{fig5}
\end{figure}

Let us hark back to the lesson from eq.~\eqref{eqn:propagated_spectrum}, that is that the propagated spectrum $\psi_j(\mathcal{R})$ is proportional to the ratio of source spectrum $q_j(\mathcal{R})$ and diffusion coefficient $\kappa(\mathcal{R})$, $\psi_j(\mathcal{R}) \propto q_j(\mathcal{R}) / \kappa(\mathcal{R})$. 
That implies that a break in the propagated spectrum can be due to a break in the source spectrum. 
Specifically, a hardening at a few hundred GV can be accommodated if the source spectrum has a hardening at this rigidity, see the top line of Fig.~\ref{fig5}.

Non-linear diffusive shock acceleration (DSA) can imprint curvature and breaks onto the source spectra through CR-induced precursors that modify the compression ratio. 
Generically, a hardening is expected below the maximum energy as low-energy particles only experience a subshock while high-energy particles feel the full compression ratio~\cite{Ptuskin:2004uk,Blasi:2005pb}. 
However, the energies and magnitudes of such features can vary with environmental conditions and source age; superposition across a realistic population tends to smear out sharp features. 
While intrinsic curvature remains plausible, it is challenging for source-only explanations to reproduce the observed sharp breaks in rigidity.

\subsection{Transport effects and breaks in \texorpdfstring{$\kappa(R)$}{kappa(R)}}

The alternative to a hardening in the source spectrum is a softening in the diffusion coefficient. 
As the diffusion coefficient enters in the denominator, this will also lead to a hardening in the propagated spectrum, see the bottom line of Fig.~\ref{fig5}. 

To understand what this could imply in terms of underlying physics, we will briefly summarise the findings from quasi-linear theory, the standard theory of wave-particle interactions~\cite{Jokipii:1966}. 
Particles interact resonantly with plasma waves. 
Specifically, the gyroradius of a particle $r_{\text{g}}$ and the wavenumber $k$ of the plasma mode satisfy the resonance condition, $r_{\text{g}} k \simeq 1$. 
This has the important consequence that if the power spectrum $W(k)$ of turbulent magnetic fields follows a power law, e.g.\ $W(k) \propto k^{-q}$, then the rigidity-dependence of the diffusion coefficient $\kappa(\mathcal{R})$ is also a power law, $\kappa(\mathcal{R}) \propto \mathcal{R}^{\delta}$ with $\delta = 2 - q$. 
Specifically, for Kolmogorov turbulence~\cite{1941DoSSR..30..301K}, $q = 5/3$ and so $\delta = 1/3$. 

If the diffusion coefficient has a break from $\kappa(\mathcal{R}) \propto \mathcal{R}^{\delta_1}$ for $\mathcal{R} \ll \mathcal{R}_b$ to $\kappa(\mathcal{R}) \propto \mathcal{R}^{\delta_2}$ for $\mathcal{R} \gg \mathcal{R}_{\text{br}}$, this would imply that the turbulent power spectrum breaks from $W(k) \propto k^{-q_2}$ for $k \ll k_{\text{br}}$ to $W(k) \propto k^{-q_1}$ for $k \gg k_{\text{br}}$; of course, $\delta_{1,2} = 2 - q_{1,2}$ and $k_{\text{br}}$ is defined such that $r_{\text{g}} k = 1$ for $\mathcal{R} = \mathcal{R}_{\text{br}}$. 
This could be realised if there were several sources of turbulence that led to different spectral shapes. 
The feature in $W(k)$ implied by a break in $\kappa(\mathcal{R})$ at hundreds of GV corresponds to wavenumbers $k$ of a few thousand $\mathrm{pc}^{-1}$, suggesting physically distinct turbulence drivers on large versus small scales. 
An example would be the presence of small-scale, self-generated turbulence dominating at large wavenumbers which are in resonance with particles of small rigidity; and external turbulence driven by stellar winds and supernova explosions dominating at small wavenumbers which are in resonance with particles of large rigidity~\cite{Evoli:2018nmb}. 

Now, which of these two options is realised in nature: intrinsic source spectral features or transport effects and breaks in $\kappa(R)$? 
A decisive discriminator between source-breaks and transport-breaks is provided by secondaries~\cite{Vladimirov:2011rn}. 
If primaries harden by $\Delta \alpha$, then in the source-break scenario, secondaries also harden by $\sim \Delta \alpha$, see top line of Fig.~\ref{fig5}. 
By contrast, if the break is in the diffusion coefficient, secondaries are affected twice because their source term is proportional to the propagated primary spectrum, and the break in secondary species is twice as strong as in primaries; see bottom line of Fig.~\ref{fig5}. 
Measurements of secondary spectral indices show a break approximately twice as large as in primaries~\cite{AMS:2018tbl}, favouring a transport-origin for the few-hundred-GV hardening.

Quantitatively fitting proton, helium, and heavier-nuclei spectra often benefits from multiple breaks in $\kappa(R)$, see, e.g., the discussion in Ref.~\cite{Schwefer:2022zly}. 
This can potentially reflect: (i) a self-generated-to-background turbulence transition, (ii) an injection/outer-scale turnover, (iii) changes in the anisotropy of turbulence, and (iv) a high-rigidity transition from resonant to small-angle, non-resonant scattering. 
The last item can steepen $\kappa(R)$ rapidly once resonant modes at the required scales are absent, providing a plausible contribution to the softening toward the knee. 

\begin{figure}
    \includegraphics[width=0.5\textwidth]{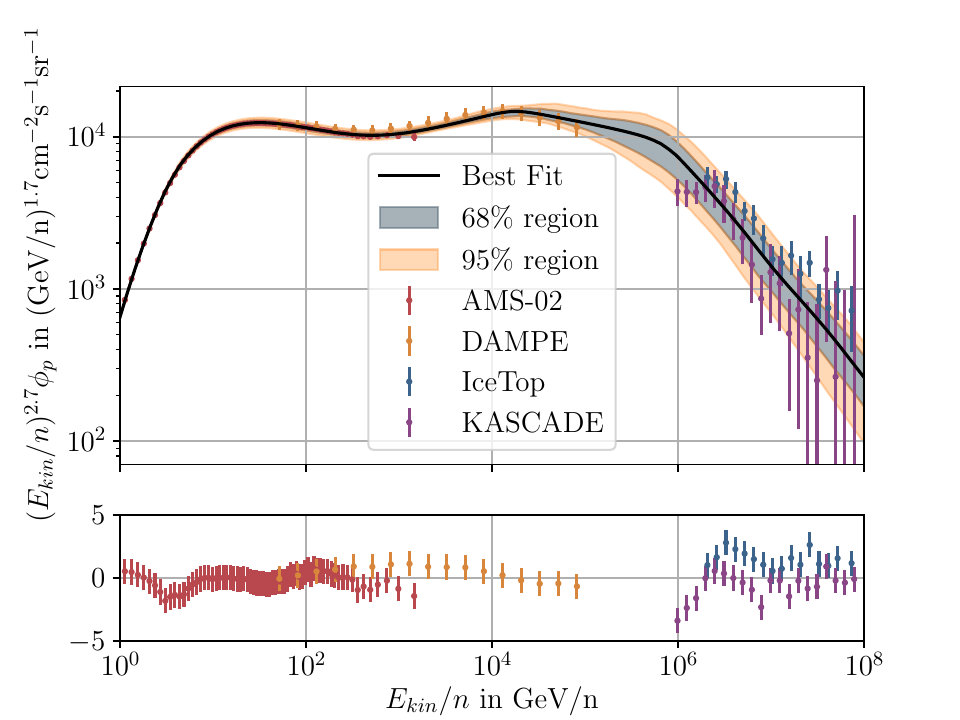}
    \includegraphics[width=0.5\textwidth]{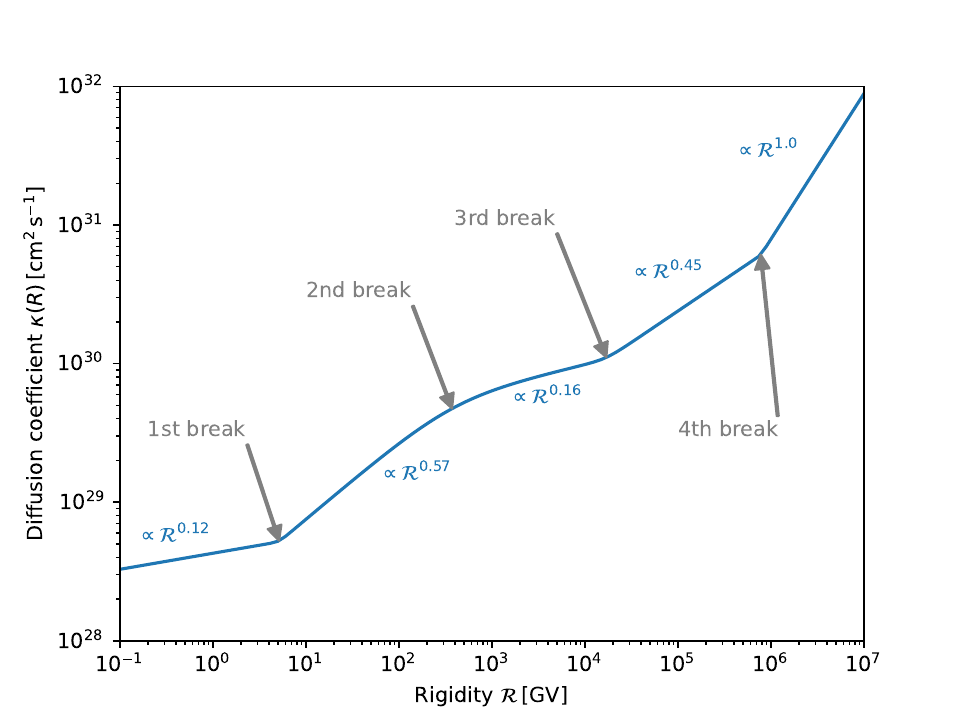}%
    \caption{
        \textbf{Left panel:} Fit to the proton spectrum over eight orders of magnitude. 
        From Ref.~\cite{Schwefer:2022zly}. 
        \textbf{Right panel:} The diffusion coefficient as a function of rigidity that is required to realise this fit. 
        }
    \label{fig6}
\end{figure}

\section{Electrons and positrons}

\begin{figure}
    \includegraphics[width=\textwidth]{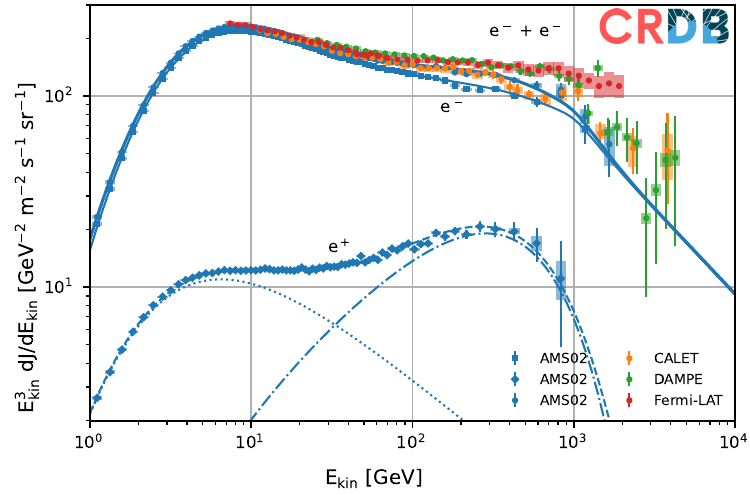}%
    \caption{Spectra of electrons and positrons, measured by the recent space-based experiments AMS-02~\cite{AMS:2021nhj}, CALET~\cite{CALET:2023emo}, DAMPE~\cite{DAMPE:2017fbg} and Fermi-LAT~\cite{Fermi-LAT:2017bpc}. Simple parametrisations for the spectra of electrons, positrons and their combination  are shown by thin solid, thin dashed and thin solid lines. 
    The contributions to the positron intensity from the secondary production process and the new, hard source of positrons is indicated by dotted and dash-dotted lines, respectively.}
    \label{fig7}
\end{figure}

In Fig.~\ref{fig7}, we show recent measurements of electron and positron fluxes. 
The data are from the space-based AMS-02~\cite{AMS:2021nhj}, CALET~\cite{CALET:2023emo}, DAMPE~\cite{DAMPE:2017fbg} and Fermi-LAT~\cite{Fermi-LAT:2017bpc}; note that there are also recent data from the ground-based H.E.S.S. observatory~\cite{HESS:2024etj}. 

The intensity of GCR positrons, shown by the diamonds, is approximately $\propto E^{-3}$ between a few GeV and a few hundred GeV. 
(Below a few GeV, solar modulation is suppressing the intensities.) 
At higher energies, the AMS-02 experiment has found the spectrum to turn over, compatible with an exponential cut-off. 
This spectrum is typically interpreted as a superposition of two contributions: 
Secondary positrons result from the production of charged pions and kaons in the inelastic interaction of nuclei with hydrogen and helium in the galactic disk. 
This is indicated by the dotted line in Fig.~\ref{fig7}. 
This contribution is significantly softer than $E^{-3}$, mostly due to cooling losses from synchrotron emission and inverse Compton scattering. 
The fact that the spectrum is much harder can then be ascribed to the presence of an additional, hard source of positrons, indicated by the dot-dashed line in Fig.~\ref{fig7}. 
The range of candidate sources is rather varied, but exotic (read: dark matter) explanations are severely constrained by gamma-ray and CMB bounds~\cite{Klasen:2015uma}. 
Possible astrophysical sources comprise pulsars/pulsar wind nebulae and old supernova remnants~\cite{Blasi:2009hv,Mertsch:2020ldv}. 

All of these sources are also sources of electrons, so they should also contribute to the GCR electron intensity and the combined electron-positron intensity. 
Measurements of these quantities are shown in Fig.~\ref{fig7} by the filled squares and circles, respectively. 
Inspection of the secondary positrons and the additional hard positron contributions allows us to estimate the corresponding electron intensities, as both are expected to be approximately charge-symmetric. 
The electron intensity (and the combined electron-positron intensity) must thus be dominated largely by the conventional source of GCR electrons, likely supernova remnants. 

The data show the measured spectrum of electrons and the combined electron-positron spectrum to approximately $\propto E^{-3.1}$ between a few GeV and about $1 \, \text{TeV}$. 
This shape can be understood as a source spectrum similar to what we assumed above for nuclei $\propto E^{-2.2}$ and softening by cooling losses during diffusion. 
Above $\sim 1 \, \text{TeV}$, the spectrum shows a clear spectral softening~\cite{DAMPE:2017fbg,CALET:2023emo,HESS:2024etj} by about one power in energy. 

The interpretation of such features is made difficult by severe energy losses of TeV electrons and positrons. 
Leptons cool rapidly via synchrotron and inverse-Compton losses, with
\begin{equation}
\frac{\dd E}{\dd t} \equiv -b(E) \simeq -b_0 E^2 \,, \quad
t_{\rm loss}(E)=\frac{1}{b_0 E}\,, \quad
\lambda(E) \simeq \sqrt{4\kappa(E)t_{\rm loss}(E)}\, . 
\end{equation}
The diffusion loss-length $\lambda(E)$ evaluates to hundreds of parsecs at TeV energies, so the highest-energy e$^{\pm}$ predominantly trace nearby, recent sources. 
Their spectra can therefore deviate from simple power laws and are sensitive to local conditions (radiation and magnetic energy densities) and to the energy scaling of $\kappa$. 
While rich in information, detailed treatment is beyond the scope of this review.

\section{Large-scale anisotropies}

Sky maps of relative intensity show dipole-like patterns with amplitudes of $\sim 10^{-4}$-$10^{-3}$ at TeV-PeV energies and a phase that notably flips by $\sim 180^\circ$ at a characteristic energy, with a dip in amplitude near the transition, e.g.~\cite{ParticleDataGroup:2024cfk}. Two principal mechanisms are relevant: 

\paragraph{Compton-Getting anisotropy.}

\begin{figure}[thp]
    \begin{minipage}[c]{.4\textwidth}
        \centering
        \includegraphics[]{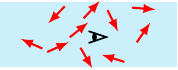}\\[2em]
        \includegraphics[]{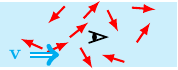}%
    \end{minipage}%
        \hfill
        \begin{minipage}[c]{.55\textwidth}
        \centering
        \includegraphics[]{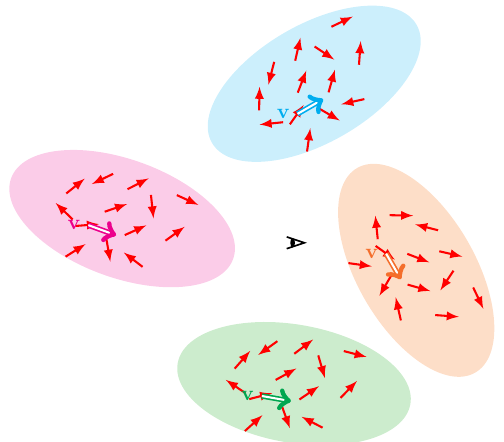}
    \end{minipage}
    \begin{minipage}[b]{.5\textwidth}
        \centering
        \caption{%
            Illustration of the conventional Compton-Getting effect. 
            \textbf{Top:} The observer is at rest with respect to the plasma hosting the turbulent magnetic field. 
            Due to frequent scattering of GCRs with turbulence, the distribution of GCRs will be isotropic. 
            \textbf{Bottom:} If the plasma has a relative velocity $\vec{v}$ with respect to the observer, the distribution of GCRs appears as boosted for the observer. 
            Expanding in powers of $\vec{v}$, the first correction is a dipolar anisotropy.
            }
    \end{minipage}%
    \hfill
    \begin{minipage}[b]{.45\textwidth}
        \centering
        \caption{%
            Illustration of the direction-dependent Compton-Getting effect. 
            The plasma in different directions around the observer is moving at different velocities, as illustrated by the different blobs. 
            In each of the blobs, the cosmic rays are isotropic in the plasma rest frame. 
            Transforming to the same observer frame requires different, direction-dependent boosts which are reflected by anisotropies on scales smaller than the dipole. 
            }
        \label{fig9}
    \end{minipage}
\end{figure}

If the observer moves at velocity $\vec{v}$ with respect to a frame in which the GCR distribution is isotropic, the observer sees an anisotropic distribution~\cite{Compton:1935wde,Gleeson:1968zza}. 
This can be computed by using the Lorentz-invariance of the phase-space density~\cite{1970P&SS...18...25F}. 
To leading order,

\begin{equation}
    \delta_{\text{CG}} \simeq \left( 2 + \frac{\partial \ln f_0}{\partial \ln p} \right) \frac{v}{c} \, , 
\end{equation}
where $f_0$ is the phase-space density in the isotropic frame and $v$ is the speed of the observer in this frame. 
The Compton-Getting effect has been observed for the Earth's orbital motion around the Sun~\cite{1986Natur.322..434C}; its Galactic-frame counterpart depends on the local standard-of-rest assumptions and is subdominant at TeV-PeV energies. 

\paragraph{Streaming anisotropy in diffusive transport.}

GCRs scatter frequently on plasma waves which results in spatial diffusion. 
If the distribution of sources is asymmetric with respect to the observer, a net flux of particles streams past them. 
Directionally, this corresponds to an anisotropy, the so-called streaming anisotropy~\cite{1966ApJ...146..480J}. 
For diffusive propagation, the dipole vector is
\begin{equation}
\vec{\delta} \simeq -\frac{3}{c} \, \kappa \cdot \vec{\nabla} \ln \psi \, , 
\end{equation}
reducing to $\vec{\delta} \simeq -(3\kappa/c) \vec{\nabla} \ln \psi$ for isotropic $\kappa$. 
Naively, model gradients often overpredict amplitudes by 1-2 orders of magnitude. 
An important correction is the presence of an ordered local magnetic field. 
With an anisotropic $\vec{\kappa}$ ($\kappa_\parallel \gg \kappa_\perp$), one has
\begin{equation}
\delta \simeq \frac{3 \kappa_{\parallel}}{c} \, \left| \bb \cdot \vec{\nabla} \ln \psi \right| \, , 
\end{equation}
where the unit vector $\bb$ indicates the direction of the regular magnetic field. 
The observable dipole is thus the projection of the true density gradient onto $\bb$, and the dipole direction aligns with $\bb$ rather than with $\vec{\nabla} \psi$. This reduces the amplitude and induces a characteristic energy dependence through the changing gradient and $\kappa_\parallel(R)$, and can produce a phase flip and amplitude dip akin to the data when a nearby young source (e.g., in the Vela region) dominates at lower energies while the large-scale gradient (e.g., toward the inner Galaxy) dominates at higher energies~\cite{Ahlers:2013ima}.

Operationally, this implies that dipole measurements constrain: (i) the local ordered field direction, (ii) which magnetic hemisphere hosts the dominant contributors at a given energy, and (iii) subdominant sources that affect $\vec{\nabla} \psi$ disproportionately to their intensity. 
High-quality measurements from HAWC, IceCube, and more recently LHAASO expand the energy lever arm for such tests.

\section{Small-scale anisotropies}

Even after subtracting low multipoles, sky maps exhibit significant small-scale structure at degree to tens-of-degrees scales with relative intensities up to $\sim 10^{-3}$. Angular power spectra display excess power up to $\ell \sim 20 \mathellipsis 50$ beyond isotropic expectations.

\begin{figure}
    \includegraphics[scale=1.2,trim={1.cm 0cm 12.5cm 0cm},clip=true]{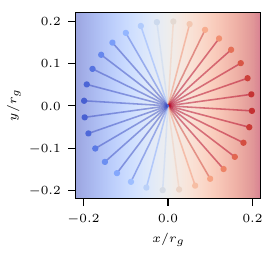} \hspace{-0.25cm}
    \includegraphics[scale=1.2,trim={1.5cm 0cm 12.5cm 0cm},clip=true]{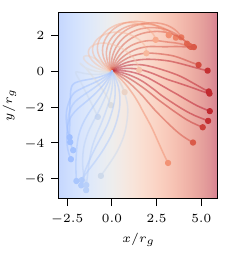} \hspace{-0.25cm}
    \includegraphics[scale=1.2,trim={1.5cm 0cm 12.5cm 0cm},clip=true]{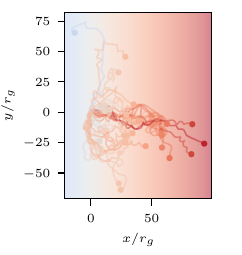} \hspace{-0.25cm}
    \caption{CR trajectories in a turbulent magnetic field, backtracked for different amounts of time $t$. \textbf{Left:} For times $t$ much smaller than the gyro time $\Omega^{-1}$, particle travel ballistically. \textbf{Middle:} At $ \sim \Omega^{-1}$ the particles are affected by the large-scale magnetic field and their trajectories start to curve. \textbf{Right:} At times $t \gg \tau_{\rm sc} \gg \Omega^{-1}$, where $\tau_{\rm sc}$ is the scattering time, particles have performed many gyrations in the average field and have been scattered by the turbulent magnetic field. It is at times $t \sim \tau_{\rm sc}$ that the turbulent magnetic field imprints itself on the small-scale anisotropies. Figure adapted from Ref.~\cite{Kuhlen:2021vvq}.}
    \label{fig8}
\end{figure}

A natural explanation is that the observer resides in a specific realization of local turbulence, and the CR arrival distribution samples the phase of turbulent modes near the last-scattering surface at a radius $\sim$ one scattering mean free path~\cite{Giacinti:2011mz,Kuhlen:2021vvq}. 
This can be understood by considering the relative diffusion of CRs in a turbulent magnetic field, see Fig.~\ref{fig8}: Whereas at early times, the particles travel balistically (left panel), the particles start gyrating in the background field at intermediate times (middle panel), before particles get scattered by the small-scale, turbulent magnetic field (right panel). 
Test-particle simulations and analytic treatments show that such mode sampling generates a multi-scale pattern with a predictable $C_\ell(\mathcal{R})$ shape, sensitive to the local scattering time $\tau_{\rm sc}(\mathcal{R})$. 
Characterising small-scale anisotropies at different rigidities therefore gives access to the local scattering time $\tau_{\rm sc}(R)$ and ultimately to turbulence properties at TV-PV rigidities where few other probes exist. 

A complementary mechanism is a generalised Compton-Getting effect in a turbulent flow, where spatially varying plasma velocities imprint small-scale modulations, see the right panel of Fig.~\ref{fig9}. 
Matching observed amplitudes requires relatively large turbulent velocity dispersions~\cite{Zhang:2024hle}; turning the argument around, small-scale anisotropies can constrain the local turbulent velocity field within a mean free path of the observer.

\section{Summary}

In this review, I summarised advances in Galactic cosmic-ray spectra and anisotropies using magnetic fields across scales as an organising principle. 
Precise, element-resolved spectra show shared rigidity breaks; I contrasted near-universal primaries with softer secondaries. 
As for the most prominent break at a few hundred GV, secondary data favour transport-driven hardening. 
The positron flux was framed as soft secondaries plus a hard component with a cut-off; I noted constraints on exotic origins and linked TeV electron softening to cooling and nearby sources. 
Compton-Getting and diffusive streaming effects were argued to give information on the magnetised medium in the galactic neighbourhood. 
Small-scale structures in the arrival direction maps arise naturally from sampling local turbulent modes near the last-scattering surface and from direction-dependent boosts, enabling constraints on $\tau_{\text{sc}}(\mathcal{R})$ and local flows. 
Overall, spectra and anisotropies jointly probe ordered and turbulent Galactic magnetic fields. 
Further progress will be guaranteed by more precise future data and refined theoretical models for transport and sources.

\bibliographystyle{JHEP}
\bibliography{ICRC2025_plenary_Mertsch}

\end{document}